\documentclass{aa}  
\usepackage{graphicx,natbib}
\bibpunct{(}{)}{;}{a}{}{,} 
\usepackage[varg]{txfonts}
\usepackage{hyperref}
\begin{document} 

   \title{Forming super-Mercuries: The role of stellar abundances}

   \author{Jingyi Mah
          \and
          Bertram Bitsch
          }

   \institute{Max-Planck-Institut f\"{u}r Astronomie, K\"{o}nigstuhl 17, 69117 Heidelberg, Germany\\
             \email{mah@mpia.de}
             }

   \date{Received ; }

 
  \abstract
  {Super-Mercuries, rocky exoplanets with bulk iron mass fraction of more than 60\%, appear to be preferentially hosted by stars with higher iron mass fraction than the Sun. It is unclear whether these iron-rich planets can form in the disc, or if giant impacts are necessary. Here we investigate the formation of super-Mercuries in their natal protoplanetary discs by taking into account their host stars' abundances (Fe, Mg, Si, S). We employ a disc evolution model which includes the growth, drift, evaporation and recondensation of pebbles to compute the pebble iron mass fraction. The recondensation of outward-drifting iron vapour near the iron evaporation front is the key mechanism that facilitates an increase in the pebble iron mass fraction. We also simulate the growth of planetary seeds around the iron evaporation front using a planet formation model which includes pebble accretion and planet migration, and compute the final composition of the planets. Our simulations are able to reproduce the observed iron compositions of the super-Mercuries provided that all the iron in the disc are locked in pure Fe grains and that the disc viscosity is low $(\alpha \sim 10^{-4})$. The combined effects of slow orbital migration of planets and long retention time of iron vapour in low-viscosity discs makes it easier to form iron-rich planets. Furthermore, we find that decreasing the stellar Mg/Si ratio results in an increase in the iron mass fraction of the planet due to a reduction in the abundance of Mg$_2$SiO$_4$, which has a very similar condensation temperature as iron, in the disc. Our results thus imply that super-Mercuries are more likely to form around stars with low Mg/Si~$(\lesssim 1)$, in agreement with observational data.}


   \titlerunning{Role of stellar abundances in the formation of super-Mercuries}
   \authorrunning{J. Mah and B. Bitsch}
   
   \maketitle
%

\section{Introduction}
The properties of the host star together with the properties of protoplanetary discs from which planets are formed are becoming necessary factors to consider for a holistic view of the process of planet formation. These information about the birth environment of planets could provide strong constraints to build better planet formation models. In particular, the elemental abundance of stars could inform (to a certain extent) the composition of its disc and potentially influence the composition of planets that emerge from the same disc \citep[e.g.][]{Bondetal2010,Marboeufetal2014,Roncoetal2015,Thiabaudetal2015,BitschBattistini2020,Jorgeetal2022}. 

For rocky planets orbiting solar-type stars, the link between stellar abundances and planetary bulk composition has been shown to exist: planetary iron mass fraction correlates with host star iron mass fraction \citep{Adibekyanetal2021sci}. Interestingly, this group of rocky exoplanets is further divided into two subgroups: super-Earths and super-Mercuries, the latter composed of $>~60~{\rm wt}\%$ iron. This suggests that the formation pathway or birth environment of the two groups could be different. 

In our Solar System, we also have an example of an iron-rich planet. Mercury's unusually massive core compared to those of the other terrestrial planets has been suggested, based on the classical view of planet formation that planets grow by collisions, to be the product of a single or multiple giant impacts which stripped away most of its primordial silicate mantle \citep[e.g.][]{Benzetal2007,AsphaugReufer2014,Chauetal2018,Jacksonetal2018,Francoetal2022}. Another model has also been put forward in which proto-Mercury's silicate mantle vapourised under high temperatures close to the Sun and most of the vapour was carried away by the gas \citep{Cameron1985}. However, Mercury's abundance of volatile elements as measured by MESSENGER \citep{Peplowskietal2011} challenges the giant impact and mantle evaporation formation scenarios due to the large amount of heating these scenarios entail.

It is unclear if super-Mercuries formed in their natal protoplanetary disc or if giant impacts played an important role their formation history. As information on the volatile abundances of super-Mercuries -- which would serve as useful constraints on their formation pathway -- are currently unavailable, it would be unfair to favour one pathway over the other. Indeed, SPH simulations of giant impacts between planetary bodies with masses of $1-20~M_{\oplus}$ show that the iron abundance of super-Mercuries can be reproduced \citep{Reinhardtetal2022}. However, the post-collision dynamical evolution of the impactor, target and fragments is not further investigated. This could indeed be an important effect because N-body simulations show that the majority of fragments might be reaccreted back onto the target itself \citep{Estevesetal2022}. 

On the other hand, there are also works that look into how the physical processes of dust growth and transport in the disc as well as the composition of the dust as a function of distance from the star could play a role in the formation of iron-rich planets without invoking giant impacts. \citet{JohansenDorn2022} show that large iron pebbles can form via nucleation from the cooling of gas in the inner regions of the disc which would go on to form iron-rich planetesimals via streaming instability. Iron-rich planets would then be a natural outcome of the accretion of iron-rich building blocks. This scenario implies that super-Mercuries should be a common occurrence as the nucleation mechanism does not depend on the host stars' or the discs' composition. Given that super-Mercuries are only found around a limited sample of stars, it is unlikely that the conditions of the discs in which they are formed are the same as those of the discs in which super-Earths – the more abundant population of the two – are formed. 

Iron-rich solids could also form around `rocklines' located in high temperature regions close to the star by means of evaporation and recondensation of dust grains, as shown by \citet{Aguichineetal2020}. They conclude that hot and massive discs are more likely to host iron-rich planets because the location of the rocklines are initially further away from the disc inner edge. The model however, is specifically tailored for the Solar System as the chemical species considered in their study are motivated by the composition of meteorites. In order to expand this model for the study of exoplanets, a more general chemistry model \citep[e.g. that suggested in][]{BitschBattistini2020,Cabraletal2023} would be more appropriate. 

Understanding the factors contributing to the formation of iron-rich solids in discs can help us identify planetary systems in which we are most likely to find super-Mercuries. In the inner Solar System, the iron fraction of planets \citep{Spohnetal2001} and meteorites \citep{TrieloffPalme2006} appear to follow a decreasing trend with increasing orbital distance. Several mechanisms that act to selectively concentrate iron into large aggregates in the inner disc have been proposed as a possible explanation \citep[e.g.][]{Wurmetal2013,Hubbard2014,KrussWurm2018}. It is unknown if discs around other stars are also iron-rich in the inner regions. Studying the formation of super-Mercuries could therefore provide further clues.

Motivated by the results of previous works, we investigate the formation of super-Mercuries via pebble accretion in their natal protoplanetary discs by additionally taking into account the host stars' abundances. Because the star and its disc form from the same material, we expect stellar abundances to play a potentially important role in influencing the outcome of planet formation because changes in stellar abundances will be reflected in the condensation sequence of different chemical species in the disc. For example, lowering the stellar Mg/Si ratio will result in more MgSiO$_3$ and SiO$_2$ to be produced instead of Mg$_2$SiO$_4$ and SiO \citep{Jorgeetal2022}. This, as we will show in this paper, has important implications for the pebble iron mass fraction and the bulk iron abundance of the planet because the condensation temperature of Mg$_2$SiO$_4$ (1354~K) is very close to that of iron (1357~K) in our model.

Our paper is structured as follows: We describe the key components of our planet formation model and the simulation set-up in Section~\ref{sec:methods}. Results from our simulations are presented in Section~\ref{sec:results}, followed by a discussion of our choice of disc parameters and other factors that could affect the pebble iron mass fraction (Section~\ref{sec:discussion}). We summarise our findings and conclude in Section~\ref{sec:conclusions}.

\section{Methods}
\label{sec:methods}
\subsection{Planet formation model}
We employed the planet formation model described in \cite{SchneiderBitsch2021a} for this work. The model includes viscous evolution of the gas disc \citep{Lynden-BellPringle1974}, pebble growth and drift \citep{Birnstieletal2012}, pebble evaporation and condensation around evaporation fronts of various chemical species in the disc \citep{SchneiderBitsch2021a}, planet growth by the accretion of pebbles \citep{JohansenLambrechts2017} and gas when the planetary seed reaches isolation mass \citep{Nduguetal2021}, and planet migration \citep{Paardekooperetal2011}. All these processes are incorporated into the \texttt{chemcomp} code \citep{SchneiderBitsch2021a} which we use for our numerical simulations. We thus only briefly describe here the important pieces relevant to our work, namely the disc structure, dust growth, pebble evaporation and recondensation, and disc composition.

\subsubsection{Gas disc model}
The gas disc model adopted is based on the classical viscous disc model \citep[e.g.,][]{Lynden-BellPringle1974}. The gas surface density profile $\Sigma_{\rm g} (r, t)$ has a power law dependence on the distance from the star and an exponential taper beyond a certain distance. Its evolution as a function of radius and time is given by \citep{Lodatoetal2017}
\begin{equation}
    \Sigma_{\rm g} (r, t) = \frac{M_{\rm disc,0}}{2\pi R_{\rm disc,0}^2}(2-\gamma)\left(\frac{r}{R_{\rm disc,0}}\right)^{-\gamma}\xi^{-\eta}\exp\left(-\frac{(r/R_{\rm disc,0})^{(2-\gamma)}}{\xi}\right),    
\end{equation}
where $M_{\rm disc,0}$ is the initial disc mass, $R_{\rm disc,0}$ is the scaling radius, $\gamma \approx 1.08$, $\eta = \frac{5/2 - \gamma}{2-\gamma}$, and $\xi = 1+\frac{t}{t_{\rm v}}$ with the viscous time $t_{\nu}$ expressed as
\begin{equation}
    t_{\nu} = \frac{R_{\rm disc,0}^2}{3(2-\gamma)^2\nu_{\rm 0}},    
\end{equation} 
where $\nu_{\rm 0}$ is the disc's kinematic viscosity evaluated at $R_{\rm disc,0}$. 

The disc's kinematic viscosity $\nu$ is parametrised by the viscosity parameter $\alpha$ \citep{ShakuraSunyaev1973} which we assumed in this work to be constant with distance from the star as well as time. The $\alpha$-viscosity parameter is related to the kinematic viscosity via
\begin{equation}
    \nu = \alpha \frac{c_{\rm s}^2}{\Omega_{\rm K}},    
\end{equation}
where $c_{\rm s}$ is the isothermal sound speed and $\Omega_{\rm K}$ the Keplerian angular velocity. 

\begin{figure}
\centering
    \resizebox{0.9\hsize}{!}{\includegraphics{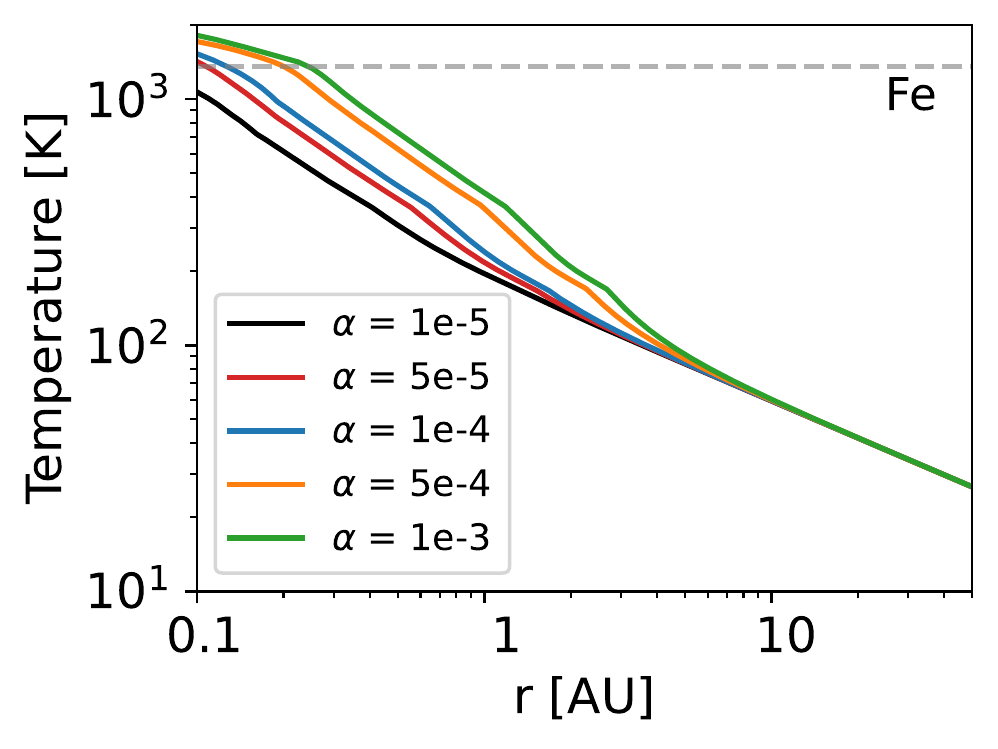}}
   \caption{Disc temperature as a function of viscosity. The higher the viscosity, the higher the temperature close to the central star due to stronger viscous heating. The evaporation temperature of iron (1357~K) is marked by the dashed line. Discs with higher viscosity are hotter and thus the evaporation front of iron (intersection between dashed line and solid lines) is located further away from the star.}
   \label{fig:disc_temp}
\end{figure}

In this work, we chose the disc mass $M_{\rm disc,0}$ to be 10\% of the stellar mass and fixed the disc radius $R_{\rm disc,0}$ to 50~AU (see also Table~\ref{tab:initial_condition} for a summary of the initial conditions). The temperature of the disc is determined by viscous heating and stellar irradiation, which dominate respectively in the inner and outer region of the disc. We show in Fig.~\ref{fig:disc_temp} the midplane temperature profiles for discs with different values of $\alpha$. Compared to low-viscosity (low $\alpha$) discs, discs with high viscosities have high temperatures in the region close to the star due to the stronger effect of viscous heating. This in turn affects the location of evaporation fronts of the chemical species in the disc. For example, the evaporation front of iron (indicated by dashed line in Fig.~\ref{fig:disc_temp}) is located further away from the star with increasing disc viscosity, and vice versa. Our disc temperature profile does not evolve with time.

\subsubsection{Dust growth}
The ratio of dust to gas in the disc is given by
\begin{equation}
    \epsilon_{\rm 0} =  \frac{\Sigma_{\rm d}}{\Sigma_{\rm g}},    
\end{equation}
where $\Sigma_{\rm d}$ is the dust surface density. The solar dust-to-gas ratio is 0.0142 \cite{Asplundetal2009}. For discs around other super-Mercury host stars, we compute $\epsilon_{\rm 0}$ by scaling it to the respective stellar iron abundances [Fe/H] using $\epsilon_{0} = 0.0142\times10^{\rm [Fe/H]}$. The dust-to-gas ratio for each star is listed in Table~\ref{tab:initial_condition}. 

A fraction of the dust in the disc is converted to pebbles following
\begin{equation}
    \Sigma_{\rm peb} = f_m \times \Sigma_{\rm d},
\end{equation}
where $f_m = 0.97$ in the drift-limited regime and $f_m = 0.75$ in the fragmentation-limited regime, following the two-population model of \citet{Birnstieletal2012}. The Stokes number of a pebble depends on its location in the disc and is computed dynamically in the simulation using
\begin{equation}
    {\rm St} = \frac{\pi}{2} \frac{R_{\rm peb}\ \rho_{\rm peb}}{\Sigma_{\rm g}},    
\end{equation}
where $R_{\rm peb}$ is the physical radius of the pebble and $\rho_{\rm peb}$ is its density. We assume $\rho_{\rm peb} = 3~{\rm g/cm^3}$ for silicate pebbles located in the region where $T > 150~K$ and 1~g/cm$^3$ for icy pebbles located in the region where $T \leq 150~K$.

\subsubsection{Pebble evaporation and recondensation}
When the pebbles drift inwards to the central star, they cross the evaporation fronts of chemical species that are present in the disc (see Table~\ref{tab:chempartition}). Two processes happen close to the evaporation fronts: First, the pebbles release the species that are locked in solid phase to the gas phase, enriching the gas with vapour of heavy elements (elements heavier than hydrogen and helium). Second, the outward diffusion of the gas allows pebbles to recondense from the vapour of heavy elements, creating localised pileups of pebbles just beyond the evaporation fronts. This mechanism is key for our work here because it enables the production of pebbles with high iron mass fraction near the iron evaporation front. When the pebbles in the pileup location drift inwards and evaporate, they further produce a large amount of iron vapour which can recondense to make even more iron-rich pebbles in a positive feedback loop. 

\subsubsection{Disc composition}
We used the chemical model based on the work of \citet{BitschBattistini2020} and assume that the initial chemical composition of the disc is the same as that of the host star. For the Sun, we adopted the elemental abundances given in \citet{Asplundetal2009} (see Table~\ref{tab:comp_sun} for the list of elements included in our model). We excluded the contribution of Na, K, Al, V, and Ti \citep[originally included in][]{SchneiderBitsch2021a} in order to make a fair comparison because the abundances of these elements are not yet available for super-Mercury host stars. The exclusion of these elements does not affect our conclusions because their abundances in the Sun are at least one order of magnitude lower compared to the other elements listed in Table~\ref{tab:comp_sun}. We also opted to exclude nitrogen because it does not play a major role in the region of the disc that is of interest here.

We next employed a simple chemical partitioning model \citep{BitschBattistini2020,SchneiderBitsch2021a} to distribute the elements in Table~\ref{tab:comp_sun} into various chemical species. The chemical species included in our model and their respective volume mixing ratios are shown in Table~\ref{tab:chempartition}. In contrast to the work of \citet{SchneiderBitsch2021a}, we make the assumption that all the iron in the disc are locked only in pure Fe grains (metallic iron) with condensation temperature of 1357~K and not partitioned into other oxidised species such as Fe$_2$O$_3$ or Fe$_3$O$_4$. This assumption is based on the results of equilibrium chemistry modelling \citep{Jorgeetal2022} and laboratory experiments \citep{Bogdanetal2023} which show that iron in the disc is present in metallic form where temperatures are higher than $\sim1000~{\rm K}$. 

In Appendix \ref{sec:appendix} we show the results of using an alternative partitioning model where the condensation temperature of enstatite (MgSiO$_3$) is assumed to be 1316~K \citep{Lodders2003} instead of 1500~K.

\begin{table}
\centering
    \caption{Solar abundances of selected elements \citep{Asplundetal2009} used in our simulations.}
    \label{tab:comp_sun}
    \begin{tabular}{c c}
    \hline\hline
    Element & Abundance\\ \hline
    He/H & 0.085 \\
    O/H & $4.90\times10^{-4}$ \\
    C/H & $2.69\times10^{-4}$ \\
    Mg/H & $3.98\times10^{-5}$ \\
    Si/H & $3.24\times10^{-5}$ \\
    Fe/H & $3.16\times10^{-5}$ \\
    S/H & $1.32\times10^{-5}$ \\ \hline
\end{tabular}
\end{table} 

\begin{table}
\centering
    \caption{Condensation temperatures \citep{Lodders2003} and volume mixing ratios \citep{BitschBattistini2020,SchneiderBitsch2021a} of chemical species included in our model.}
    \label{tab:chempartition}  
    \begin{tabular}{c c c}
    \hline\hline
    Species & $T_{\text{cond}}$~(K) & volume mixing ratio \\ \hline
    CO            & 20  & 0.45 $\times$ C/H  \\
    CH$_4$        & 30  & 0.45 $\times$ C/H  \\
    CO$_2$        & 70  & 0.1  $\times$ C/H \\
    H$_2$S        & 150 & S/H  \\
    H$_2$O        & 150 & O/H - (3 $\times$ MgSiO$_3$/H + \\
           &      & 4 $\times$ Mg$_2$SiO$_4$/H + CO/H + \\
           &      & 2 $\times$ CO$_2$/H) \\
    Mg$_2$SiO$_4$ & 1354 & Mg/H - Si/H \\
    Fe            & 1357 & Fe/H \\
    MgSiO$_3$     & 1500 & Mg/H - 2 $\times$ (Mg/H - Si/H) \\ \hline
    \end{tabular}
\end{table} 

\subsection{Initial conditions}
We ran six sets of simulations in total to model discs around the Sun and five super-Mercury host stars (K2-38, K2-106, K2-229, Kepler-107 and Kepler-406). As these stars have different masses, we chose -- for consistency's sake -- the stellar luminosity to be the value at 1~Myr (the time when planet formation is assumed to take place) from the stellar evolutionary model of \citet{Baraffeetal2015}. The disc mass is fixed at 10\% the stellar mass and the disc size is set to 50~AU. The inner and outer edges of the disc are located at 0.1~AU and 1000~AU, respectively, with the simulation region divided into 500 cells on a logarithmically-spaced grid.

The dust-to-gas ratio in the disc around each super Mercury host star is scaled according to their respective [Fe/H] value with the reference value for the Sun taken from the protosolar value listed in \citet{Asplundetal2009}. The initial composition of each disc is then computed according to the measured abundances of Fe, Mg and Si of the host star \citep{Adibekyanetal2021sci}. For the abundance of sulphur, we assumed that they scale the same way as silicon \citep{Chenetal2002} and set [S/H] = [Si/H]. Table~\ref{tab:initial_condition} shows a summary of our initial conditions.

For each star, we tested for six different values of the alpha-viscosity parameter $\alpha = 10^{-5}, 5\times10^{-5}, 10^{-4}, 2\times10^{-4}, 5\times10^{-4}, 10^{-3}$ and three values of pebble fragmentation velocity $u_{\rm frag} = 1, 5, 10~{\rm m/s}$. The combinations of these two free parameters give 18 simulations for each stellar system. All simulations were ran for 3~Myr (the assumed lifetime of our gas disc) with a time step of 10~yr.

\begin{table*}
\centering
    \caption{Initial conditions of our simulations.}
    \label{tab:initial_condition}
    \begin{tabular}{c c c c c c c c c c}
    \hline\hline
    Star      & $M_*~[M_{\odot}]$ & $L_*~[L_{\odot}]$ & $M_{\rm disc,0}~[M_{\odot}]$ & $R_{\rm disc,0}~[{\rm AU}]$ & $\epsilon_{\rm 0}$ & [Fe/H] & [Mg/H] & [Si/H] & [S/H]\\ \hline
    Sun        & 1.00 & 1.93 & 0.100 & 50 & 0.0142 & 0     & 0     & 0     & 0 \\
    K2-38      & 1.07 & 2.16 & 0.107 & 50 & 0.0258 & 0.26  & 0.24  & 0.27  & 0.27 \\
    K2-106     & 0.95 & 1.78 & 0.095 & 50 & 0.0179 & 0.10  & 0.07  & 0.05  & 0.05 \\
    K2-229     & 0.84 & 1.46 & 0.084 & 50 & 0.0124 & -0.06 & -0.07 & -0.06 & -0.06 \\
    Kepler-107 & 1.24 & 2.74 & 0.124 & 50 & 0.0373 & 0.42  & 0.41  & 0.37  & 0.37 \\
    Kepler-406 & 1.07 & 2.16 & 0.107 & 50 & 0.0247 & 0.24  & 0.23  & 0.25  & 0.25 \\\hline
\end{tabular}
\end{table*} 

\section{Results}
\label{sec:results}
\subsection{Pebble iron mass fraction}
\begin{figure*}
\centering
   \resizebox{\hsize}{!}{\includegraphics{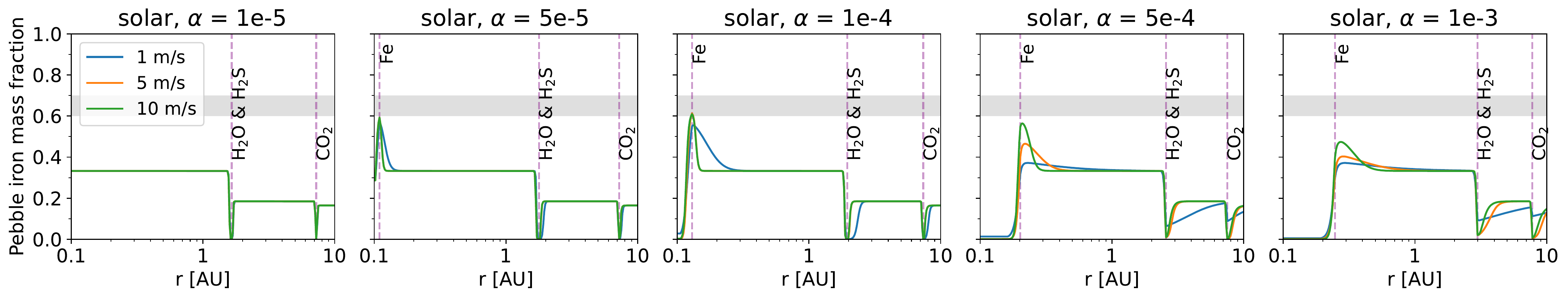}}
   \caption{Iron mass fraction of pebbles in the disc around the Sun as a function of disc viscosity and pebble fragmentation velocity. Purple dashed lines mark the location of the evaporation fronts of the various chemical species included in our model. Grey bands represent the range of values for the iron abundance of Mercury \citep{Bruggeretal2018,Adibekyanetal2021sci}. Increasing the disc's viscosity results in a lower pebble iron mass fraction because the rapid inflow of gas from the outer disc continuously transports freshly-produced iron vapour away.}
   \label{fig:disc_ironfrac}
\end{figure*}

Fig.~\ref{fig:disc_ironfrac} shows the iron mass fraction of pebbles in the disc around the Sun as a function of disc viscosity and the pebble fragmentation velocity. We computed this quantity by dividing the surface density of iron in each cell with the surface density of the pebbles in the same cell. We show here the results for the disc around the Sun as an example. The results for discs around other super-Mercury host stars are largely similar but with different peak values for the pebble iron mass fraction.

The pebble iron mass fraction is low in the region far away from the star because volatile species such as CH$_4$, CO$_2$ and H$_2$O, which are the most abundant species in our model, are solid in the colder outer disc and thus make up most of the mass of the pebbles. Closer to the central star these volatile species evaporate and are removed from the pebbles, resulting in an increased contribution of iron to the total mass of the pebbles.

The pebbles are the most iron-rich around the iron evaporation front due to the recondensation of iron vapour to make new iron-rich pebbles at this location. By the same reasoning, the iron mass fraction near the evaporation fronts of the volatile species drops sharply because there is an increase in the amount of pebbles that condensed from the vapour of these volatile species. In the region within the iron evaporation front the pebble iron mass fraction drops to zero because all the iron in the pebbles has evaporated.

Varying the disc viscosity $\alpha$ affects the location of the iron evaporation front (see Fig.~\ref{fig:disc_ironfrac}) as well as the maximum value of the pebble iron mass fraction near the evaporation front. An increase in the value of $\alpha$ means an increase in the strength of viscous heating and a higher temperature in the inner disc which pushes the evaporation fronts of iron and water to larger orbital distances (the location of the CO$_2$ evaporation front is not strongly affected because it is located in the region where the disc is heated mainly by stellar irradiation). In the example shown here, the iron evaporation front is absent when $\alpha = 10^{-5}$ because the disc is too cold -- the iron evaporation front is located closer to the star than the disc's inner edge; the iron evaporation front starts to appear and moves further outwards as we increase the value of $\alpha$ (see also Fig.~\ref{fig:disc_temp}).

However, increasing the disc viscosity has a detrimental effect on the pebble iron mass fraction near the iron evaporation front. In high-viscosity discs the timescale of viscous evolution is shorter, which translates into iron vapour being carried away quickly along with the gas onto the star. This reduces the amount of iron vapour that can diffuse outwards and recondense.

Increasing the pebble fragmentation velocity $u_{\rm frag}$ helps to increase the pebble iron mass fraction in discs with $\alpha \geq 10^{-4}$. This is because the pebbles can grow to a larger size and therefore release a larger amount of iron vapour to the gas when they evaporate, making available more iron vapour for recondensation. For lower viscosity discs, the effect of increasing $u_{\rm frag}$ is not very pronounced because the pebbles in a low-viscosity disc are already large, regardless of the value of the fragmentation velocity \citep[in the fragmentation regime the Stokes number ${\rm St} \propto u_{\rm frag}^2/\alpha$;][]{Birnstieletal2009}. 

For the example that we show here, we find that the pebble iron mass fraction computed from our model is barely consistent with the iron abundance of Mercury (grey bands in Fig.~\ref{fig:disc_ironfrac}); we discuss Mercury's formation in Section~\ref{sec:mercury}. Given the elemental abundances of the Sun we use in our simulations (Table~\ref{tab:comp_sun}), the maximum pebble iron mass fraction we can achieve with our model, which includes pebble evaporation and recondensation, is $\sim60\%$ in the best case scenario with $\alpha = 10^{-4}$. This is in fact related to the relative abundance of magnesium and silicon with respect to iron, which we discuss in the following subsection. Nevertheless, we can already conclude from this subsection that a disc with low-viscosity (but also sufficiently hot that the iron evaporation front is present!) is most favourable for the growth of iron-rich planets because it is able to retain iron vapour for a long period of time.

\subsection{Influence of stellar abundances on pebble iron mass fraction}
\begin{figure}
\centering
   \resizebox{0.9\hsize}{!}{\includegraphics{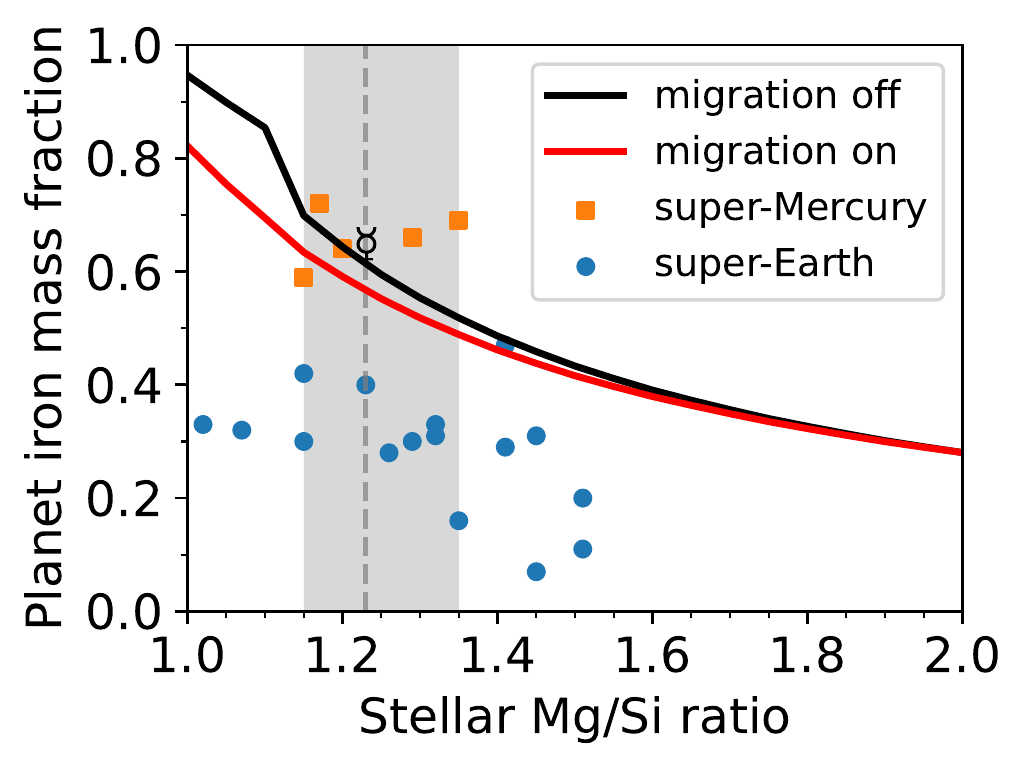}}
   \caption{Maximum iron mass fraction of a planet formed near the iron evaporation front as a function of the Mg/Si ratio of the central star. These are results from simulations using solar disc composition, $\alpha = 10^{-4}$ and $u_{\rm frag} = 5~{\rm m/s}$. The iron mass fraction of super-Mercuries and super-Earths \citep{Adibekyanetal2021sci} as well as Mercury are included for comparison. The grey band and dashed line represent the range of Mg/Si values of super-Mercury host stars and the Sun, respectively.}
   \label{fig:planetFe_stellarMgSi}
\end{figure}

Here we report how the elemental abundances of the host star affect the amount of iron that is available in the disc. Among the chemical species included in our model, iron and Mg$_2$SiO$_4$ have very similar condensation temperatures $(1357~{\rm K~vs.~} 1354~{\rm K})$. As we have briefly mentioned in the Introduction, the stellar Mg/Si ratio determines how much Mg$_2$SiO$_4$ is present relative to MgSiO$_3$ in the disc, with Mg$_2$SiO$_4$ becoming the major condensate as the stellar Mg/Si ratio increases \citep{Jorgeetal2022}. This is also reflected in our partitioning model (Table~\ref{tab:chempartition}). The model first computes the amount of Mg$_2$SiO$_4$ from the total amount of Mg and Si available in the disc before allocating the remaining amount of Mg and Si to MgSiO$_3$. For the solar Mg/Si value, the ratio of Mg$_2$SiO$_4$ to MgSiO$_3$ in the disc is $\approx 0.3$. This ratio increases when the stellar Mg/Si ratio increases and vice versa. 

Due to the very similar condensation temperatures of Mg$_2$SiO$_4$ and iron, an increase in the abundance of Mg$_2$SiO$_4$ in the disc with respect to iron translates into a lowered contribution of iron to the total mass of the pebbles. The stellar Mg/Si ratio, which determines the relative abundance of the aforementioned chemical species, therefore becomes an important factor as it determines the maximum amount of iron that can be incorporated into the pebbles (and subsequently, planets) under favourable disc conditions (i.e. low viscosity). 

Consequently, we investigate the relationship between the iron mass fraction of the planets and the host star's Mg/Si ratio. We set up a disc with $\alpha = 10^{-4}$, $u_{\rm frag} = 5~{\rm m/s}$, solar initial disc composition, and placed a planetary seed around the iron evaporation front $(0.12~{\rm AU} \leq a \leq 0.16~{\rm AU})$. We compared results between the simulation in which the planetary seed migrate as it grows and the simulation in which it does not. We repeated the experiment 20 times with different values of semi-major axis $a$ and recorded the maximum value for the planet iron mass fraction. We then changed the stellar Mg abundance and repeated the whole experiment. The results are shown in Fig.~\ref{fig:planetFe_stellarMgSi}.

As expected, the planet iron mass fraction decreases as we increase the Mg abundance (and therefore the Mg/Si ratio) of the host star. With more Mg available in the disc, the pebbles now contain more Mg$_2$SiO$_4$ which in turn lowers the mass fraction of Fe. Activating disc-induced orbital migration in the simulations returns in a reduction in the final iron mass fraction of the planet due to the accretion of iron-poor pebbles interior to the iron evaporation front as the planets migrate towards the star (we elaborate on this in the next subsection). 

We also plot the iron mass fraction of super-Mercuries and super-Earths listed in \citet{Adibekyanetal2021sci} in Fig.~\ref{fig:planetFe_stellarMgSi} for comparison. While there is a spread in the stellar Mg/Si ratio for super-Earth host stars, the range of stellar Mg/Si ratios appear rather restricted for super-Mercury host stars (grey band in figure), in support of our simulation results which suggest that it could be difficult to form these iron-rich planets in discs around stars with high Mg/Si ratio.

\subsection{Conducive conditions for super-Mercury formation}
\begin{figure*}
\centering
   \resizebox{0.9\hsize}{!}{\includegraphics{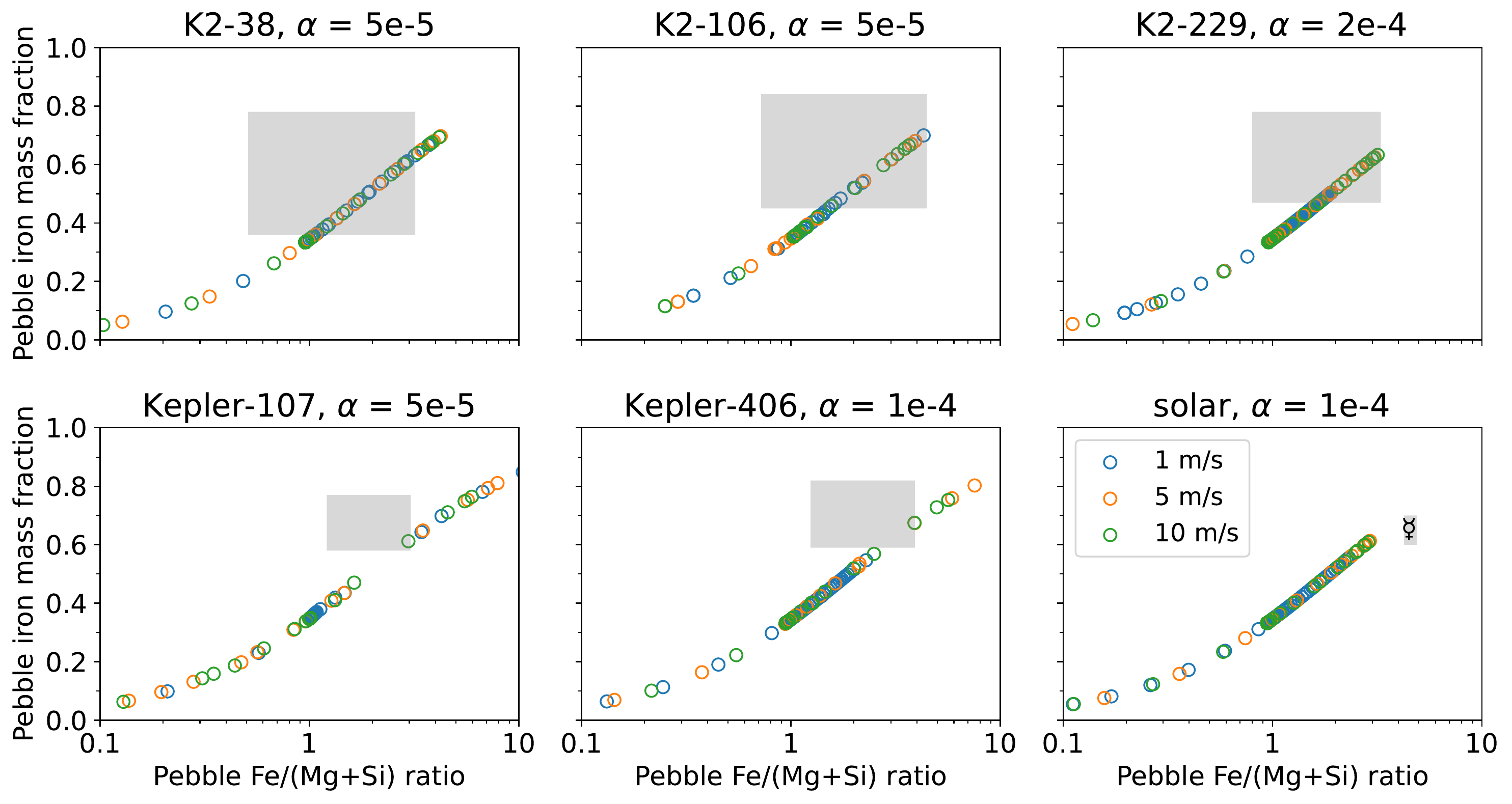}}
   \caption{Iron mass fraction versus iron-to-silicate ratio of pebbles in discs around different super-Mercury host stars (see Table~\ref{tab:initial_condition}). Shown here are the best outcome for each system with the optimum value of $\alpha$. Grey boxes represent the range of possible compositions of the observed super-Mercuries in these systems \citep{Adibekyanetal2021sci}.}
   \label{fig:ironfrac_FeMgSi}
\end{figure*}

\begin{figure*}
\centering
   \resizebox{0.9\hsize}{!}{\includegraphics{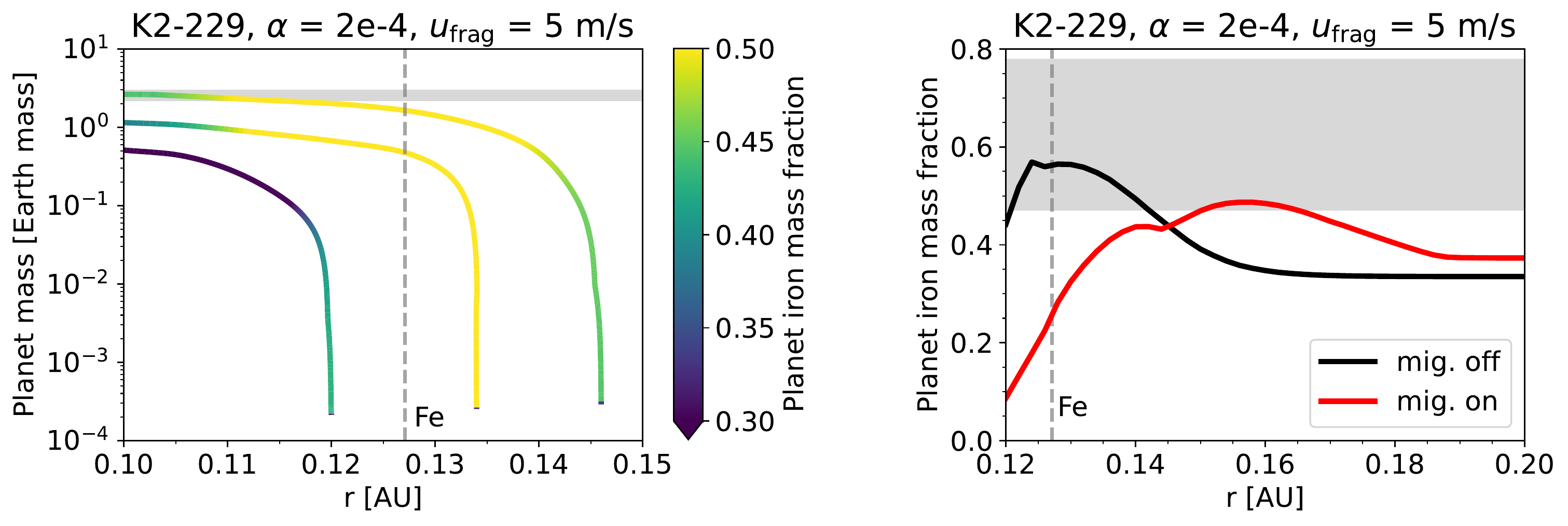}}
   \caption{Planetary growth tracks and iron mass fraction as a function of orbital distance from the star. {\it Left panel:} Example growth tracks of planetary seeds placed initially in the vicinity of the iron evaporation front (grey dashed line) and the evolution of their iron mass fraction. Planetary seeds accrete iron-rich pebbles as they approach the iron evaporation front while the planetary iron mass fraction starts to drop as they migrate past the iron evaporation fornt. The horizontal grey band encompass the mass range of K2-229b. {\it Right panel:} Final iron abundance of a migrating and non-migrating planet as a function of their initial semi-major axis. The maximum iron mass fraction achieved by a migrating planet is lower because it accretes iron-poor material as it migrates toward the disc inner edge. The grey band represents the range of values for the iron mass fraction of K2-229b.}
   \label{fig:growthtrack_K2-229}
\end{figure*}

We next report the overall performance of our planet formation model when applied to super-Mercury systems. In addition to their iron mass fraction, we also have constraints on the iron-to-silicate ratio (Fe/(Mg+Si)) of the super-Mercuries from \citet{Adibekyanetal2021sci} which we can use to evaluate the performance of our model. We scanned through our simulation results (similar to Fig.~\ref{fig:disc_ironfrac}) for all super-Mercury systems and selected for the best case scenario which we show in Fig.~\ref{fig:ironfrac_FeMgSi}. We defined `best case scenario' as the simulation output which the pebble iron mass fraction around the iron evaporation line is maximum. 

We find that our model predictions are consistent with the compositions of the super-Mercuries when the disc viscosity is low. Low-viscosity discs are conducive environments for the formation of iron-rich pebbles because they allow for larger grains and have slower viscous evolution timescales, both which aid in keeping iron vapour around the iron evaporation front for a longer period. Our simulation results show that $\alpha \leq 2\times10^{-4}$ works best for all the systems we have investigated in this work. Although for some super-Mercury host stars the optimum value of $\alpha$ can be as low as $5\times10^{-5}$, we understand that $\alpha$ values lower than $10^{-4}$ are difficult to achieve in real discs \citep[e.g.][]{Dullemondetal2018,Flahertyetal2018}.

\subsection{Case study: Formation of K2-229 b}
Finally, we show results of simulations to form a super-Mercury in the disc around the star K2-229 (Fig.~\ref{fig:growthtrack_K2-229}). We set $\alpha = 2\times10^{-4}$ (the optimum value for this system derived from earlier results) and $u_{\rm frag} = 5~{\rm m/s}$. We then placed a planetary seed in the disc between $0.12~{\rm AU} \leq a \leq 0.20~{\rm AU}$ and tracked its mass and composition with time. The initial mass of the planetary seed corresponds to the mass when pebble accretion becomes efficient. The experiment was repeated 20 times with different values of $a$ but within the same range.

The left panel of Fig.~\ref{fig:growthtrack_K2-229} shows the growth tracks of three selected planetary seeds. The initial iron mass fraction of the planetary seed depends on their initial semi-major axis. As the planet moves towards the iron evaporation front (dashed line in figure), it accretes iron-rich pebbles and become more iron-rich; further inward migration results in the accretion of iron-poor pebbles which lowers the planet's iron mass fraction. 

The right panel of Fig.~\ref{fig:growthtrack_K2-229} shows a comparison between the results with and without orbital migration. With migration, as we have discussed in the previous paragraph, planets initially located further away from the iron evaporation front end up with the highest iron mass fraction. The maximum iron mass fraction that can be reached is also lower as the planet inevitably accretes iron-poor pebbles when it crosses the iron evaporation front. If migration is excluded, then planets initially close to the evaporation front are the most iron-rich. This is expected because the pebble iron mass fraction is the highest at this location. Accretion of iron-rich pebbles naturally produce iron-rich planets. Taken together, our results suggest that the most conducive initial condition to form iron-rich planets is to have the planetary seeds start close to the iron evaporation front in a low-viscosity disc. Planets can then take advantage of long orbital migration timescale and viscous evolution timescale to accrete as many iron-rich pebbles as possible, which are only present in a narrow region of the disc.

\section{Discussion}
\label{sec:discussion}
\subsection{Disc parameters}
In this work we investigated the disc parameters that support the formation of iron-rich pebbles and sustain them in the disc for a sufficiently long period of time for iron-rich planets to form. We found that discs with low viscosity $(\alpha \sim 10^{-4})$ are preferred due to their long viscous evolution timescales and the slow migration speed of planets. The influence of pebble fragmentation velocity is negligible when $\alpha$ is low because grain size is already large in low-viscosity discs.

While discs with $\alpha = 10^{-4}$ are consistent with observations \citep[see][]{Flahertyetal2018,Dullemondetal2018}, the iron evaporation front (located at 0.15~AU for a $0.1~M_{\odot}$ disc around a $1~M_{\odot}$ star) in these discs could be very close to the disc inner edge where the magnetorotational instability (MRI) is active \citep[see][for a review]{Lesuretal2022}. As disc viscosity is ramped up in the MRI-active region, the growth of dust grains in this region could possibly be inhibited, posing potental problems for the growth of pebbles (and planets). On the other hand, it has been showed that large bodies could still grow near the silicate sublimation line at $\sim 0.1~{\rm AU}$ \citep{Flocketal2019} or within the `Curie line' at $T \sim 1000~{\rm K}$ \citep{Bogdanetal2023}. Detailed modelling of disc temperature and density profile in the inner region of protoplanetary discs is beyond the scope of this paper, but would be needed in future studies to understand the mechanism of dust growth in regions very close to the star. 

Our results also show high disc viscosities do not work in favour of forming super-Mercuries because iron vapour is swiftly transported away with the gas. This reduces the enrichment of iron in the disc that is generated by the pebble recondensation mechanism. Nevertheless, if it is indeed the case that the inner regions of discs are highly viscous or turbulent environments with $\alpha = 10^{-3}$, our results show that the pebble iron mass fraction can still be as high as about 40\% (higher than the stellar iron abundance) when the pebble fragmentation velocity is 10~m/s. In the context of our model, the formation of iron-rich planets in highly viscous discs would then involve collisions, but the number of collisions needed will be reduced as the solids in the disc are already enriched in iron.

\subsection{Solar abundances and the formation of Mercury}
\label{sec:mercury}
Since our model predicts that pebbles in the solar protoplanetary disc contain at most 60\% iron, the process of pebble accretion alone cannot explain the formation of Mercury. Other additional mechanisms such as giant impacts would be needed to help boost the iron mass fraction by mantle stripping \citep[e.g.][]{Benzetal2007,AsphaugReufer2014,Jacksonetal2018,Francoetal2022}. 

Invoking other physical processes is however, not the only way to increase the iron mass fraction of the solids in the disc. We show in this work that the pebble iron mass fraction also depends on the host star's abundances, especially the relative abundance of Fe versus Mg and Si. Here we used the solar elemental abundances from \cite{Asplundetal2009} for which the solar Mg/Si ratio is computed to be 1.23. If this ratio is lesser, then it would mean that there is more iron in the disc. To reach pebble iron mass fraction of 70\%, the solar Mg/Si ratio would need to be at about 1.10 according to our simulation results (Fig.~\ref{fig:planetFe_stellarMgSi}). 

The elemental abundances of the Sun have been recently revisited, with the solar Mg/Si ratio revised down to 1.10 \citep{Asplundetal2021} and 0.91 \citep{Maggetal2022}. With these revised values, we expect pebbles with iron mass fraction comparable to the composition of Mercury to form in the solar protoplanetary disc and the issue of insufficient iron in the disc would be mitigated. Furthermore, since the stellar abundances are computed using the solar value as reference, we also expect an increased pebble iron mass fraction in discs around the other super-Mercury host stars, which could make our model predictions better comport with the compositions of Kepler-107c and Kepler-406b.

\subsection{Planet formation model}
The planet formation model we use here considers that planets grow by the accretion of pebbles. As our focus in this work lies in understanding where and how iron-rich pebbles can form in discs around solar type stars, we did not explore the effects of planet growth via planetesimal accretion. It is likely that planetesimals, formed by the streaming instability mechanism, were also present in the protoplanetary disc, although the formation efficiency of planetesimals and their exact spatial distribution in the disc are not well constrained.

We do not expect drastic differences in the iron mass fraction of planets located close to the star if they were to additionally accrete planetesimals. This is because the composition of the planetesimals also follows the disc composition (they grow from the pebbles in the disc) and since iron remains in the solid phase in most regions of the disc due to its high condensation temperature, the amount of iron contained in the solids (whether pebbles or planetesimals) is expected to be the same everywhere beyond the iron evaporation front. Furthermore, if the mechanisms to form iron-rich solids in the inner disc (as mentioned in the introduction) are indeed acting, then planetesimals which form in the disc would be iron-rich and the accretion of the iron-rich planetesimals would further increase the final iron mass fraction of the planet.

In this work we have assumed a constant disc temperature profile for simplicity. As a consequence, iron-rich solids are found only in a narrow region of the disc. More realistically, the evaporation fronts in the disc would move with time in response to changes in disc heating sources (e.g. stellar luminosity). The region where iron-rich solids form would then extend over a wider region, potentially creating an iron gradient as in the Solar System -- a conjecture that would be interesting to test. 

We have also neglected the evolution of the system when the gas disc has dissipated. The iron-rich solids and leftover solid material would likely undergo a stage of giant impacts to form larger iron-rich bodies that are part of a system with multiple planets \citep{Bonomoetal2019}. Studying the exact formation pathway of specific super-Mercury systems would require additional N-body simulations, an endeavour which we reserve for the future. 

The scarcity of super-Mercuries, could be an indication that specific conditions, both of the star and the disc, are required for their formation. Our approach suggests that the stellar Mg/Si ratio could be an important parameter to consider going forward as it determines the maximum pebble iron mass fraction that can be achieved given a specific set of disc parameters. Our model also predicts that there should be no super-Mercuries around stars with high Mg/Si ratios. Super-Mercuries could still form in discs which do not fulfil the aforementioned conditions, in which case other mechanisms could be at work.

\section{Conclusions}
\label{sec:conclusions}
We have studied the possibility of forming super-Mercuries in discs around stars of different abundances using a coupled disc evolution and planet formation model that takes into account pebble evaporation and recondensation. This is done by exploring the influence of disc viscosity $(10^{-5} \leq \alpha \leq 10^{-3})$, pebble fragmentation velocity $(u_{\rm frag} = 1, 5, 10~{\rm m/s})$, as well as the stellar Mg/Si ratio on the pebble iron mass fraction. We found that super-Mercuries are most likely to form in low-viscosity discs around stars with low Mg/Si ratio. The long viscous evolution timescale allows iron vapour to remain in the disc for a long enough period while the long orbital migration timescale allows planetary seeds to migrate slowly and maximise the accretion of iron-rich pebbles. The pebble iron mass fraction predicted by our model is consistent with the compositions of the observed super-Mercuries but is barely consistent with Mercury's composition, suggesting that collisions or other iron-concentrating mechanisms in the disc likely play a role in the latter's formation. Finally, we demonstrated that the production of iron-rich pebbles and larger bodies which contribute to the building blocks of iron-rich planets can be made easier by considering the process of pebble evaporation and recondensation around the iron evaporation front.

\begin{acknowledgements}
      We thank Gerhard Wurm for the insightful comments and suggestions that helped to improve the manuscript. J.M. and B.B. acknowledge the support of the DFG priority program SPP 1992 ``Exploring the Diversity of Extrasolar Planets'' (BI 1880/3-1). B.B. thanks the European Research Council (ERC Starting Grant 757448-PAMDORA) for their financial support. 
\end{acknowledgements}

\bibliographystyle{aa} 
\bibliography{main} 

\begin{appendix} 
\section{An alternative disc chemical partitioning model}
\label{sec:appendix}
Because we excluded elements that less abundant than sulphur (such as Na, K, Al, Ti and V) in this alternative, chemical species that are more refractory than iron are absent. Iron therefore becomes the last chemical species to evaporate when we assume 1316~K for the condensation temperature of MgSiO$_3$ (instead of 1500~K as in the main paper). In such a model, solids located between the evaporation front of MgSi$_2$O$_4$ and iron are therefore composed of pure metallic iron, and there would be no solid material in the region within the iron evaporation front because all solids are evaporated and the dust surface density will go to zero (in theory). 

\begin{figure}
\centering
   \resizebox{\hsize}{!}{\includegraphics{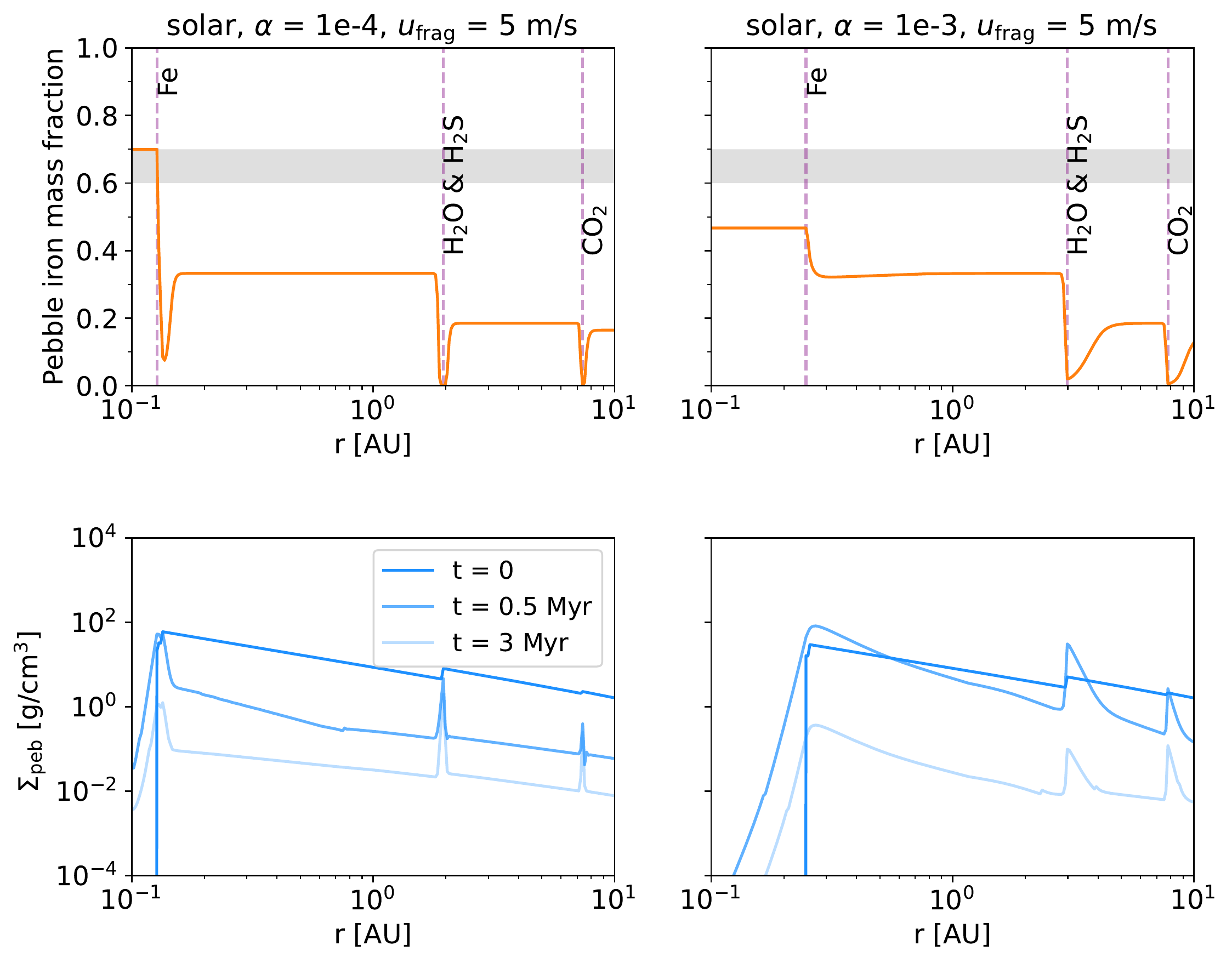}}
   \caption{Pebble iron mass fraction as a function of orbital distance and time evolution of pebble surface density profile for two example discs with solar composition and $\alpha = 10^{-4}, 10^{-3}$. This is the outcome when we employ a chemical partitioning model where $T_{\rm cond} = 1316~{\rm K}$ for MgSiO$_3$.}
   \label{fig:disc_ironfrac_alternative}
\end{figure}

We show in Fig.~\ref{fig:disc_ironfrac_alternative} how the pebble surface density profile changes with time and the corresponding pebble iron mass fraction as a function of orbital distance, for two examples of disc viscosity. At $t = 0$ the distribution of solids in the disc shows a cutoff at the location of the iron evaporation front due to the exclusion of other chemical species with higher condensation temperature than iron. At later times the distribution of solids near the iron evaporation front become more spread out but the overall amount of solids remains very low. Consequently, our code returns a constant value for the pebble iron mass fraction everywhere within the iron evaporation front. 

In this model, iron-rich pebbles form exactly (and only) at the location of the iron evaporation front. The conditions to form of iron-rich planets becomes even more restricted: planets must form exactly at the iron evaporation front without migrating, in order to achieve a high iron mass fraction. 

\end{appendix}

\end{document}